\documentclass[twocolumn]{revtex4-2}
\usepackage[utf8]{inputenc}
\usepackage[english]{babel}
\pdfoutput=1 
\usepackage{amsmath}
\usepackage{subcaption}
\usepackage{esint}
\usepackage{pgfplots}
\usepackage[margin=2.5 cm]{geometry}
\pgfplotsset{
    compat=newest
}
\usepackage{physics}
\usepackage{graphicx}
\usepackage{mathtools}
\usepackage{wrapfig}
\usepackage{tikz}
\usepackage{etoolbox}
\usetikzlibrary{shapes.geometric}
\usepackage{hyperref}

\begin{document}

\title{The Anatomy of a Topological Phase Transition in a 2D Chern Insulator}
\author{Arjo Dasgupta, Indra Dasgupta}

\affiliation{School of Physical Sciences, \\ Indian Association for the Cultivation of Science, Kolkata}

\begin{abstract}
The onset of the topological phase transition in a two-dimensional model for a Chern Insulator, namely the Qi-Wu-Zhang(QWZ) model, is illustrated, with particular emphasis on the appearance of {\it chiral edge-modes}. The edge-modes are studied by analysing the dynamics of the edge-states in an equivalent model for a  one-dimensional charge pump, using a technique known as {\it dimensional extension}. A further {\it real-space analysis} allows us to explain the onset of the topological phase transition in terms of time-reversal symmetry breaking and to quantitatively study the localisation of the edge-modes.
\end{abstract}

\maketitle

\section{Introduction} 
Chern insulators are two-dimensional materials which break time-reversal symmetry and are characterised by a {\it Quantised Hall Conductance} and the existence of {\it Chiral Edge Modes}. Both these properties can be linked to the {\it nonzero Chern number} calculated over the Brillouin Zone, as in the case of the Integer Quantum Hall Effect. They can thus be understood as exhibiting a ``Hall effect without a magnetic field'', as in the case of the model suggested by Haldane \cite{P10}. 

A topological phase transition in a Chern insulator is characterised by a change in the Chern number of the system which occurs on continuously tuning a parameter across some critical value. Phases corresponding to different Chern numbers are topologically distinct and cannot be connected via adiabatic transformations \cite{P6}. A topological phase transition can be detected by the appearance of {\it robust} chiral edge modes and a jump in the Hall conductance. 

This paper provides a pedagogical introduction to this phenomenon.
Our aim is to {\it explain} the appearance of chiral edge-modes and to provide a mechanism for the onset of the topological phase transition, in a model for a 2D Chern insulator, namely the Qi-Wu-Zhang(QWZ) model \cite{P2}. 

We approach the question of the chiral edge modes by first considering the Su-Schrieffer-Heeger (SSH) model, a 1-dimensional lattice model with staggered hopping amplitudes, which is known to host {\it topologically protected edge-states}. On varying the parameters of this model adiabatically, the edge-states acquire a dynamics, and by making this variation cyclic in time we arrive at the Rice-Mele (RM) model for {\it adiabatic charge pumping} which is defined on the $(k,t)$ space. 

Following \cite{P6}, we present a technique for analysing the dynamics of edge modes in the charge pump using a simple {\it toy model}, which can be deformed into the original model. 

The topological features of the RM model are determined by the Chern number calculated over $(k,t)$ space. By tuning the hopping strengths in the toy-model, we demonstrate a topological phase transition in the RM model, which leads to a change in the quantity of charge pumped per cycle.

We then construct the QWZ model from the RM model by transforming the cyclic time-coordinate to a second momentum variable $(t \rightarrow k_y)$. This method, called {\it dimensional extension} allows us to analyse the chiral edge modes which are created in the topological phase transition, in terms of the dynamics of the edge states of the charge pump.

In order to provide a physical mechanism for the onset of the topological phase transition, we study the real-space properties of the QWZ model and separate the terms in the model into an onsite potential, and a hopping term which breaks time-reversal symmetry. It is found that the parameter $\lambda$ which controls the topological phase transition in the RM model also controls the term which breaks time-reversal symmetry in the QWZ model. On tuning $\lambda$ beyond a critical value $\lambda_c$ at which the bulk-gap closes, the system passes from the trivial phase with Chern number $n=0$ to the nontrivial phase with $n=1$. 

The real-space Hamiltonian also allows us to explore the formation of edge-states numerically. We find that the edge-states {\it separate out} from the bulk eigenstates as the system moves into the nontrivial phase. We also find a limit of the model in which the edge-modes are restricted to  lattice site at the two edges.  The formation of these {\it exactly-localised} edge-modes are also demonstrated numerically.

The dimensional extension procedure as presented in \cite{P6} provides a connection between the topological phase transitions in the charge pump and in the model for the Chern insulator. Our analysis goes further to suggest a {\it physical mechanism} which explains the onset of the topological phase transition in terms of {\it time-reversal symmetry breaking}, and makes use of the real-space properties to demonstrate the formation of edge states. 
 
This paper is organised as follows. The Su-Schrieffer-Heeger model, a prototype for the 1D charge pump is introduced in Section \ref{sshmodel}, and both its bulk properties and its topologically protected edge-states are studied. The bulk features of the Rice-Mele charge pump and the dynamics of its edge-states are studied in Section \ref{rmmodel}. In Section \ref{qwzmodel}, the Qi-Wu-Zhang model is constructed using dimensional extension and the bulk-boundary correspondence for this model is illustrated. Finally, the mechanism for the topological phase transition is presented in Section \ref{mechanism}, followed by conclusions in Section \ref{conc}.

\section{The Su-Schrieffer-Heeger Model: Bulk-Edge Correspondence}\label{sshmodel}

The Su-Schrieffer-Heeger(SSH)\cite{P11} model is a tight-binding model defined on a 1D lattice with two distinct lattice sites per unit cell (see Fig. 1), and staggered hopping amplitudes, ($w$ for intercell hopping, $v$ for intracell hopping).

\begin{figure}[h]
\centering
 \begin{tikzpicture}[scale=0.7]
  \node at (1.4,-0.45) {$w$};
  \node at (2.4,-0.15) {$v$};
  
  \draw [thick,red] (-2,-0.5)--(-1,0);
  \draw [thick,green] (-1,0) -- (0,-0.5);
  \draw [thick,red] (0,-0.5)--(1,0);
  \draw [thick,green] (1,0) -- (2,-0.5);
  \draw [thick,red] (2,-0.5) -- (3,0);
  \draw [thick,green] (3,0) -- (4,-0.5);
  \draw [thick,red] (4,-0.5)--(5,0);
  \draw [thick,green] (5,0)--(6,-0.5);
  \draw [thick,red] (6,-0.5)--(7,0);
  \node at (-1,0.25) {$b$};
  \node at (-2.1,-0.75) {$a$};
 
  \draw [fill,black](-2,-0.5) circle (0.075cm);
  \draw [fill,black](0,-0.5) circle (0.075cm);
  \draw [fill,black](2,-0.5) circle (0.075cm);
  \draw [fill,black](4,-0.5) circle (0.075cm);\draw [fill,black](6,-0.5) circle (0.075cm);
  \draw [fill,gray](-1,0) circle (0.075cm);
  \draw [fill,gray](1,0) circle (0.075cm);
  \draw [fill,gray](3,0) circle (0.075cm);
  \draw [fill,gray](5,0) circle (0.075cm);
  \draw [fill,gray](7,0) circle (0.075cm);

 \end{tikzpicture}
 \caption{\small SSH Model: 1D chain with staggered hopping amplitudes. Neighbouring sites form different sublattices $a$ and $b$}
\end{figure}
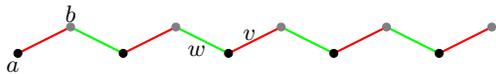

\begin{eqnarray}\label{SSH_Ham}
\nonumber
 H_{SSH}= &v& \sum_{m=1}^{N} (b^\dagger_m a_m+hc)\\
 &+&w\sum_{m=1}^{N-1}(a^\dagger_{m+1}b_m+hc)
\end{eqnarray}

Crucially, hopping necessarily changes the sublattice index ($a/b$) of the state. 

To study the bulk properties of this system, we assume that the edges have no effect on the bulk physics. This allows us to enforce {\it periodic boundary conditions}.

We can write the Hamiltonian in terms of the Fourier components of the field operators $a_k,b_k$ as $H_{SSH} = \sum_k (a^\dagger_k \>\> b^\dagger_ k) H(k) (a_k \>\> b_k)^T$ where the {\it bulk momentum-space Hamiltonian} $H(k)$ is given by:

\begin{equation}\label{SSH_k}
  H(k) =
  \begin{bmatrix}
   0 & v+we^{-ika} \\
   v+we^{ika} & 0
  \end{bmatrix}
\end{equation}

The dispersion relation can be read off from the eigenvalues of $H(k)$ as  

$E_{\pm}(k)=\pm \sqrt{v^2+w^2+2vw\cos(ka)}$. 

\begin{figure}[h]
\centering
\includegraphics[width=\columnwidth]{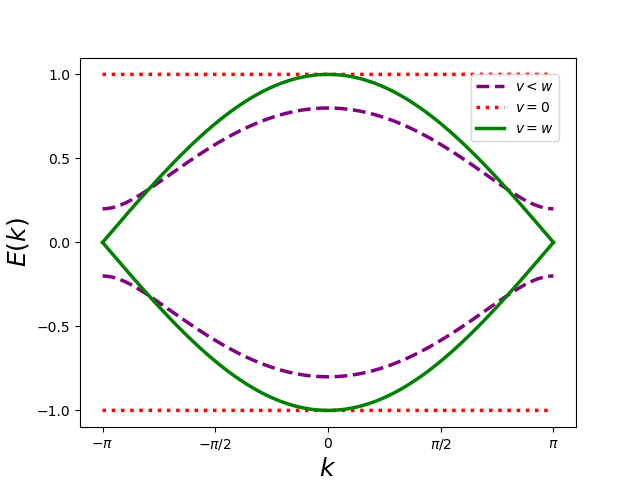}
 \caption{\small Dispersion relations of the SSH Model. For $v=w$ the system is metallic while for $v=0$ (or $w=0$) the bands are flat.}
\end{figure}

The minimum gap between the two bands is thus $\Delta_{min} = 2|v-w|$. For the case of equal hopping strengths ($v=w$), the bands touch at the edge of the Brillouin Zone $k=\pm\frac{\pi}{a}$ and the system is metallic. In general, $v\neq w$, and the system is gapped (see Fig. 2 for details).

For $v=0$ or $w=0$, {\it the bands are flat} ($\frac{\partial E }{\partial k}=0$). This tells us that as the SSH chain breaks apart into dimers, the group velocity of electrons on the chain vanishes: there is no electron-transport in the bulk.

We note that the bulk Hamiltonian can be written as $H(k) = {\bf d}(k).\vec \sigma$ with ${\bf d}(k)=(v+w\cos(ka),w\sin(ka),0)$.

We can thus think of the Hamiltonian as a map from the Brillouin Zone $S^1$  to the parameter space $R^2 \setminus \{0\}$. In the SSH model, ${\bf d}(k)$ describes a circle in the punctured plane, which encloses the origin if $v<w$. For a two-dimensional parameter space, the Berry phase of the system is a topological invariant quantised as $2\pi n$ where $n$ is the winding number of the curve in $R^2 \setminus \{0\}$. Thus we have two topologically distinct phases: $n=0$ for $v>w$ (trivial), and $n=1$ for $v<w$ (nontrivial).

In order to determine the experimental consequences of the topology of the system, we need to place the SSH model on an {\it open chain}.

To start with, we study the two flat-band limits: $(v=1,w=0)$ (trivial) and $(v=0,w=1)$ (nontrivial).

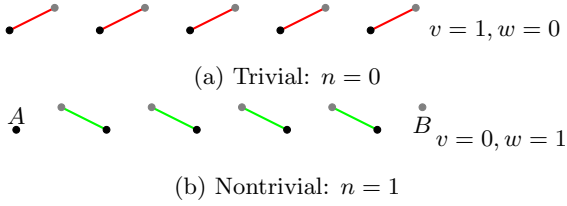
\begin{figure}[h]
\centering
\begin{subfigure}{\columnwidth}
 \begin{tikzpicture}[scale=0.6]
 \node at (8.75,-0.5) {$v=1,w=0$};
  \draw [thick,red] (-2,-0.5)--(-1,0);
  \draw [thick,red] (0,-0.5)--(1,0);
  \draw [thick,red] (2,-0.5) -- (3,0);
  \draw [thick,red] (4,-0.5)--(5,0);
  \draw [thick,red] (6,-0.5)--(7,0);
  \draw [fill,black](-2,-0.5) circle (0.075cm);
  \draw [fill,black](0,-0.5) circle (0.075cm);
  \draw [fill,black](2,-0.5) circle (0.075cm);
  \draw [fill,black](4,-0.5) circle (0.075cm);
  \draw [fill,black](6,-0.5) circle (0.075cm);
  \draw [fill,gray](-1,0) circle (0.075cm);
  \draw [fill,gray](1,0) circle (0.075cm);
  \draw [fill,gray](3,0) circle (0.075cm);
  \draw [fill,gray](5,0) circle (0.075cm);
  \draw [fill,gray](7,0) circle (0.075cm);
 \end{tikzpicture}
 \caption{Trivial: $n=0$}
 \label{triv}
\end{subfigure}
\begin{subfigure}{\columnwidth}
 \begin{tikzpicture}[scale=0.6]
  
  \node at (8.75,-2.75) {$v=0,w=1$};
  \draw [thick,green] (-1,-2) -- (0,-2.5);
  \draw [thick,green] (1,-2) -- (2,-2.5);
  \draw [thick,green] (3,-2) -- (4,-2.5);
  \draw [thick,green] (5,-2)--(6,-2.5);
  \node at (7,-2.4) {$B$};
  \node at (-2,-2.2) {$A$};
  \draw [fill,black](-2,-2.5) circle (0.075cm);
  \draw [fill,black](0,-2.5) circle (0.075cm);
  \draw [fill,black](2,-2.5) circle (0.075cm);
  \draw [fill,black](4,-2.5) circle (0.075cm);
  \draw [fill,black](6,-2.5) circle (0.075cm);
  \draw [fill,gray](-1,-2) circle (0.075cm);
  \draw [fill,gray](1,-2) circle (0.075cm);
  \draw [fill,gray](3,-2) circle (0.075cm);
  \draw [fill,gray](5,-2) circle (0.075cm);
  \draw [fill,gray](7,-2) circle (0.075cm);
 \end{tikzpicture}
 \caption{Nontrivial: $n=1$}
 \label{ntriv}
\end{subfigure}
 \caption{\small Flat-band limits. The trivial limit in (a) does not host edge states, while the nontrivial limit in (b) hosts zero energy edge-states.}
\end{figure}

In addition to the bulk eigenstates with energy $\pm 1$, the nontrivial phase (Fig. 3(b)) also allows for the existence of zero energy eigenstates confined to the edges: $H\ket{1,A} = H\ket{N,B}=0$. This comes about because $[H,a^\dagger_1]=[H,b^\dagger_N]=0$.

This fact suggests a connection between the topological phase of the bulk of the system and the existence of such edge-states, an idea which which we shall explore further in other contexts. For a rigorous discussion on the bulk-edge correspondence in the SSH model we refer the reader to Chapter 1 of \cite{P6}.

\section{The Rice-Mele Model: Adiabatic Charge Pumping}\label{rmmodel}

The modern theory of polarisation \cite{P3} links the change in the electric polarisation in a one-dimensional crystal during a {\it cyclic, adiabatic} change in a parameter $\lambda$ to the Berry Phase calculated over the $(k,\lambda)$ space: 

\begin{equation}
 \Delta P = \frac{e}{2\pi} \oint dk \oint d\lambda \grad \cross \bra{u_k}\ket{\grad u_k}
\end{equation}

Where the states $\ket{u_k}$ are the Bloch eigenstates of $H(k)=E(k)\ket{u_k}$, with corresponding wavefunctions $u_k(r)$ which are periodic in the lattice-sites $u_k(r+a)=u_k(r)$. 

This fact allows us to construct an {\it adiabatic charge pump}: by varying the parameters of the Hamiltonian for a 1D crystal such that it {\it switches} between a topologically trivial and a nontrivial phase, we can get a change in the electric polarisation in the wire, i.e., a quantity of charge is {\it pumped across the wire}. 

The amount of charge pumped in the cycle is given by:

\begin{equation}\label{Pump}
\Delta P = ne 
\end{equation}

where $n$ is the Chern number \cite{P3} calculated over the $(k,\lambda)$ space. 

Adiabatic charge pumping on a SSH chain may be implemented by adding an onsite potential $u$ and varying the parameters cyclically in time:

\begin{eqnarray}\label{RiceMele}
\nonumber
 H(t)&=& v(t) \sum_{m=1}^{N} (b^\dagger_m a_m+hc)\\
 &+&w\sum_{m=1}^{N-1}(a^\dagger_{m+1}b_m+hc) \\
 \nonumber
 &+& u(t)\sum_{m=1}^{N}(a^{\dagger}_m a_m - b^{\dagger}_m b_m)
\end{eqnarray}

with: $u(t)=\lambda \sin(\frac{2\pi t}{T})$, $v(t)= 1 + \lambda \cos(\frac{2\pi t}{T})$,$w(t)=\lambda$, where $\lambda$ is a parameter which varies from $-1$ to $+1$. This is called the Rice-Mele(RM) model for adiabatic charge pumping.
 
In momentum space, we have:

\begin{equation*}
 H(k,t) = 
 \begin{bmatrix}
   u(t) & v(t)+we^{-ika} \\
   v(t)+we^{ika} & -u(t)
  \end{bmatrix} 
\end{equation*}
 
 The parameters ${\bf d}(k,t)$ of the model Hamiltonian $H(k,t) = {\bf d}(k,t).\vec \sigma$ trace out a torus in the 3D parameter space: 
 
 \begin{eqnarray*}
  d_x(k,t)&=&1+\lambda \cos(\frac{2\pi t}{T})+\lambda \cos(ka)\\
  d_y(k,t)&=&\lambda \sin(ka)\\
  d_z(k,t)&=&\lambda \sin(\frac{2\pi t}{T})
 \end{eqnarray*}
 
Suppose $\lambda>0$. At $t=0$, the parameters sweep out the circle $(1+\lambda(\cos(ka)+1),\lambda\sin(ka),0)$, which clearly misses the origin: $n=0$, the system is in the trivial phase. 
 
At $t=\frac{T}{2}$, the parameters are $(1+\lambda(\cos(ka)-1),\lambda \sin(ka),0)$. Thus if $\lambda>\frac{1}{2}$ the system reaches the nontrivial phase at $t=\frac{T}{2}$ before coming back to the trivial phase. 

Thus at $\lambda=\lambda_c=\frac{1}{2}$ we have a topological phase transition from a trivial phase with $n=0$ to a nontrivial phase with $n=1$.

For negative values of $\lambda$ we obtain a similar phase transition from $n=0$ to $n=-1$ at $\lambda_c = -\frac{1}{2}$. Thus for $|\lambda| > \frac{1}{2}$, one unit of  charge is pumped across the chain in one cycle $\Delta P = e$. The sign of $\lambda$ determines the direction of pumping.  

In order to obtain further insights on the dynamics of the edge states in the Rice-Mele model, we deform it into a simpler toy model as suggested in \cite{P6}, in which the parameters are varied in a piecewise linear fashion:

\begin{equation*}
\begin{matrix}
    &  & 8t & t<\frac{1}{8}\\
    &  & 1 &\frac{1}{8}\leq t <\frac{1}{4}\\
    u(t)& = & 4(\frac{1}{2}-t) &\frac{1}{4}\leq t <\frac{3}{4}\\
    &  & -1 &\frac{3}{4}\leq t < \frac{7}{8}\\
    &  & 8(t-1) & \frac{7}{8}\leq t \leq 1   
\end{matrix} 
\end{equation*}
\begin{equation}\label{ToySeq}
\begin{matrix}
    &  & 2 & t<\frac{1}{16}\\
    &  &  3+ \lambda - 16(\lambda+1)t&\frac{1}{16}\leq t <\frac{1}{8}\\
    v(t)& = & 1-\lambda &\frac{1}{8}\leq t <\frac{7}{8}\\
    &  & -13 - 15 \lambda + 16(\lambda+1)t &\frac{7}{8}\leq t < \frac{15}{16}\\
    &  & 2 & \frac{15}{16}\leq t \leq 1   
\end{matrix}
\end{equation}

\begin{equation*}
\begin{matrix}
    &  & 0 & t<\frac{1}{8}\\
    &  & 8\lambda(t-\frac{1}{8}) &\frac{1}{8}\leq t <\frac{1}{4}\\
    w(t)& = & \lambda &\frac{1}{4}\leq t <\frac{3}{4}\\
    &  & 8\lambda(\frac{7}{8}-t)&\frac{3}{4}\leq t < \frac{7}{8}\\
    &  & 0 & \frac{7}{8}\leq t \leq 1   
\end{matrix}  
\end{equation*}

For the toy model defined above with $\lambda > \frac{1}{2}$, the parameters ${\bf d}(k,t)$ map out a surface which is topologically equivalent to a torus enclosing the origin, and thus the system has a Chern number of 1 (see Fig. 4(a)), while for $\lambda<\frac{1}{2}$ the surface fails to enclose the origin, and the Chern number vanishes (see Fig. 4(b)).

\begin{figure}[h]
\centering
\begin{subfigure}{\columnwidth}
\begin{tikzpicture}[scale=0.8]
\node at (-0.5,1.5) {$\lambda=1$};
\node at (-0.5,-0.25) {$d_x$};
\draw [stealth-] (-0.5,0)--(-0.75,0); 
\node at (-1.25,1.3) {$d_y$};
\draw [stealth-] (-1.5,1.5)--(-1.8,1.2); 
\draw [stealth-] (-3,2.75)--(-3,2);
\node at (-2.7,3) {$d_z$};
\node at (-2.4,2.5) {$t=\frac{1}{8}$};
\node at (-2.4,-2.7) {$t=\frac{7}{8}$};
\draw (-5.5,0)--(-0.5,0);
\draw (-4.5,-1.5)--(-1.5,1.5);
\draw(-3.0,-2)--(-3.0,2);
\draw [red,thick,dashed] (-3.0,0) ellipse (1.0 and 0.4);
\draw [red,thick,dashed] (-3.0,2) ellipse (1.0 and 0.4);
\draw [red,thick,dashed] (-3.0,-2) ellipse (1.0 and 0.4);
\draw [gray,thick] (-1,0)--(-1,2);
\draw [red,fill] (-1,2) circle (0.05cm);
\draw [gray,thick] (-1,2)--(-3,2);
\draw [cyan,thick,stealth-] (-3.5,2)--(-3,2);
\draw [cyan,thick,stealth-] (-2.5,2)--(-3,2);
\draw [cyan,thick,stealth-] (-3.25,1.75)--(-3,2);
\draw [cyan,thick,stealth-] (-2.75,2.25)--(-3,2);
\draw [gray,thick,dashed] (-3.0,1) ellipse (1.0 and 0.4);
\draw [gray,thick,dashed] (-3.0,-1) ellipse (1.0 and 0.4);
\draw [cyan,thick,stealth-] (-3,1.1)--(-3,1.5);
\draw [cyan,thick,stealth-] (-3,.1)--(-3,.5);
\draw [cyan,thick,stealth-] (-3,-.9)--(-3,-.5);
\draw [cyan,thick,stealth-] (-3,-1.9)--(-3,-1.5);

\draw [gray,thick] (-3,-2)--(-1,-2);
\draw [gray,thick] (-1,-2)--(-1,0);
\draw [cyan,thick,stealth-] (-1,1.2)--(-1,0.8);
\draw [cyan,thick,stealth-] (-1,-0.8)--(-1,-1.2);
\draw [cyan,thick,stealth-] (-1.9,2)--(-1.5,2);
\draw [cyan,thick,stealth-] (-1.5,-2)--(-1.9,-2);
\node at (-4.5,2) {$t=\frac{1}{4}$};
\node at (-4.5,-2) {$t=\frac{3}{4}$};
\draw [fill,red] (-3,2) circle (0.05cm);
\draw [fill,red] (-1,-2) circle (0.05cm);
\draw [fill,red](-1,0) circle (0.05cm);
\node at (-0.4,0.25) {$t=0$};
\draw [cyan,thick,stealth-] (-3,-2)--(-3.5,-2);
\draw [cyan,thick,stealth-] (-3,-2)--(-2.5,-2);
\draw [cyan,thick,stealth-] (-3,-2)--(-3.25,-2.25);
\draw [cyan,thick,stealth-] (-3,-2)--(-2.75,-1.75);
\draw [fill,red] (-3,-2) circle (0.05cm);
\node at (-4.5,0.3) {$t=\frac{1}{2}$};
\draw [fill] (-3,0) circle(0.05cm);
\end{tikzpicture} 
\caption{$\lambda = 1$: $n=1$}
\end{subfigure}
\begin{subfigure}{\columnwidth}
\begin{tikzpicture}[scale=0.8]
\node at (-0.2,1.5) {$\lambda=0.3$};
\node at (-0.5,-0.25) {$d_x$};
\draw [stealth-] (-0.5,0)--(-0.75,0); 
\node at (-1.25,1.3) {$d_y$};
\draw [stealth-] (-1.5,1.5)--(-1.8,1.2); 
\draw [stealth-] (-3,2.75)--(-3,2);
\node at (-2.7,2.6) {$d_z$};
\draw (-5.5,0)--(-0.5,0);
\draw (-4.5,-1.5)--(-1.5,1.5);
\draw(-2.3,-2)--(-2.3,2);
\draw(-3,-2)--(-3,2);
\draw [red,thick,dashed] (-2.3,0) ellipse (0.3 and 0.12);
\draw [red,thick,dashed] (-2.3,2) ellipse (0.3 and 0.12);
\draw [red,thick,dashed] (-2.3,-2) ellipse (0.3 and 0.12);
\draw [gray,thick] (-1,0)--(-1,2);
\draw [red,fill] (-1,2) circle (0.05cm);
\draw [gray,thick] (-1,2)--(-2.3,2);
\draw [gray,thick,dashed] (-2.3,1) ellipse (.3 and 0.12);
\draw [gray,thick,dashed] (-2.3,-1) ellipse (.3 and 0.12);
\draw [cyan,thick,stealth-] (-2.3,1.1)--(-2.3,1.5);
\draw [cyan,thick,stealth-] (-2.3,.1)--(-2.3,.5);
\draw [cyan,thick,stealth-] (-2.3,-.9)--(-2.3,-.5);
\draw [cyan,thick,stealth-] (-2.3,-1.9)--(-2.3,-1.5);
\draw [gray,thick] (-2.3,-2)--(-1,-2);
\draw [gray,thick] (-1,-2)--(-1,0);
\draw [cyan,thick,stealth-] (-1,1.2)--(-1,0.8);
\draw [cyan,thick,stealth-] (-1,-0.8)--(-1,-1.2);
\draw [cyan,thick,stealth-] (-1.9,2)--(-1.5,2);
\draw [cyan,thick,stealth-] (-1.5,-2)--(-1.9,-2);
\node at (-3.5,2) {$t=\frac{1}{4}$};
\node at (-3.5,-2) {$t=\frac{3}{4}$};
\draw [fill,red] (-2.3,2) circle (0.05cm);
\draw [fill,red] (-1,-2) circle (0.05cm);
\draw [fill,red](-1,0) circle (0.05cm);
\node at (-0.25,0.25) {$t=0$};

\draw [fill,red] (-2.3,-2) circle (0.05cm);
\node at (-3.5,0.3) {$t=\frac{1}{2}$};
\draw [fill] (-3,0) circle(0.05cm);
\end{tikzpicture} 
\caption{$\lambda = 0.3$: $n=0$}
\end{subfigure}
\caption{\small Nontrivial and trivial phases of the charge pump. For $\lambda>0.5$ (a) the parameters $(d_x,d_y,d_z)$ trace out a torus enclosing the origin and $n=1$ , while for $\lambda<0.5$ (b) the torus misses the origin and thus $n=0$.} 
\end{figure}
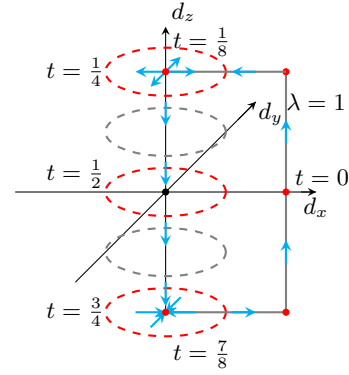
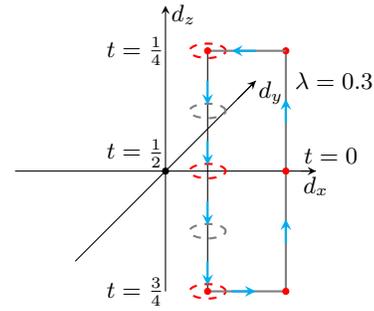

We are particularly interested in the behaviour of those instantaneous eigenstates of the toy model sequence \ref{ToySeq} with $\lambda=1$ (nontrivial), which may be localised to the left or right edge. Recall that, for fixed $v$, the appearance of edge-states is controlled by the intercell hopping strength $w$. For $0\leq t<\frac{1}{8}$ and $\frac{7}{8} \leq t <1$,  $w=0$, there is no intercell hopping, and edge states are forbidden entirely (see Fig. 5(a)). Edge states at the left and right end begin to appear as $t$ crosses $\frac{1}{8}$, and $w$ is turned on.

During the time interval from $t=\frac{1}{4}$ to $t=\frac{3}{4}$ we have $w(t)=1,v(t)=0$ and $u(t)$ decreases linearly from $+1$ to $-1$. The situation is similar to the topologically nontrivial phase of the SSH model in the fully dimerised limit, (see Fig. 5(b)). 

\begin{eqnarray}\label{ham}
\nonumber
 H(t)&=&\sum_{m=1}^{N-1}(a^\dagger_{m+1}b_m+hc) \\&+& u(t)\sum_{m=1}^{N}(a^{\dagger}_m a_m - b^{\dagger}_m b_m)
\end{eqnarray}

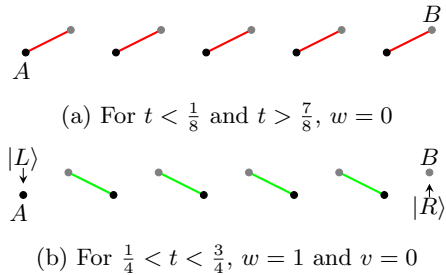
\begin{figure}[h]
\centering
\begin{subfigure}{\columnwidth}
 \begin{tikzpicture}[scale=0.6]
  \draw [thick,red] (-1,0) -- (-2,-0.5);
  \draw [thick,red] (1,0) -- (0,-0.5);
  \draw [thick,red] (3,0) -- (2,-0.5);
  \draw [thick,red] (5,0)--(4,-0.5);
  \draw [thick,red] (7,0)--(6,-0.5);
  \node at (7,0.35) {$B$};
  \node at (-2.1,-0.9) {$A$};
 
  \draw [fill,black](-2,-0.5) circle (0.075cm);
  \draw [fill,black](0,-0.5) circle (0.075cm);
  \draw [fill,black](2,-0.5) circle (0.075cm);
  \draw [fill,black](4,-0.5) circle (0.075cm);
  \draw [fill,black](6,-0.5) circle (0.075cm);
  \draw [fill,gray](-1,0) circle (0.075cm);
  \draw [fill,gray](1,0) circle (0.075cm);
  \draw [fill,gray](3,0) circle (0.075cm);
  \draw [fill,gray](5,0) circle (0.075cm);
  \draw [fill,gray](7,0) circle (0.075cm);
 \end{tikzpicture}
 \caption{For $t<\frac{1}{8}$ and $t>\frac{7}{8}$, $w=0$}
\end{subfigure}
\begin{subfigure}{\columnwidth}
 \begin{tikzpicture}[scale=0.6]  
  \draw [thick,green] (-1,0) -- (0,-0.5);
  \draw [thick,green] (1,0) -- (2,-0.5);
  \draw [thick,green] (3,0) -- (4,-0.5);
  \draw [thick,green] (5,0)--(6,-0.5);
  \node at (7,0.35) {$B$};
  \node at (-2.1,-0.9) {$A$};
  \draw [fill,black](-2,-0.5) circle (0.075cm);
  \draw [fill,black](0,-0.5) circle (0.075cm);
  \draw [fill,black](2,-0.5) circle (0.075cm);
  \draw [fill,black](4,-0.5) circle (0.075cm);
  \draw [fill,black](6,-0.5) circle (0.075cm);
  \draw [fill,gray](-1,0) circle (0.075cm);
  \draw [fill,gray](1,0) circle (0.075cm);
  \draw [fill,gray](3,0) circle (0.075cm);
  \draw [fill,gray](5,0) circle (0.075cm);
  \draw [fill,gray](7,0) circle (0.075cm);
  \node at (-2,0.3) {$\ket{L}$};
  \draw [stealth-] (-2,-0.25)--(-2,0.1); 
  \draw [stealth-] (7,-0.25)--(7,-0.55); 
  \node at (7,-0.8) {$\ket{R}$};
 \end{tikzpicture}
 \caption{For $\frac{1}{4}<t<\frac{3}{4}$, $w=1$ and $v=0$}
\end{subfigure}
\caption{\small Before $t=\frac{1}{8}$ no edge states appear (a). From $t=\frac{1}{4}$ to $t=\frac{3}{4}$ (b), the right edge-state is pumped up in energy while the left edge-state is pumped down}
\end{figure}

Consider the left and right edge-states $\ket{L}=\ket{1,A}$ and $\ket{R}=\ket{N,B}$. From \ref{ham} we find that: $H(t)\ket{L}=u(t)\ket{L}$ and $H(t)\ket{R}=-u(t)\ket{R}$. As $u$ goes from $1$ to $-1$, the right edge state is pumped up in energy from the lower to the upper band while the left edge state is pumped down in energy. 

As demonstrated in \cite{P6}, the number of edge states pumped up  per cycle is a topological invariant. The Rice-Mele model which is topologically equivalent to the toy model with $\lambda>\frac{1}{2}$ thus also has a  right edge-state which is pumped up in energy $\left(\frac{dE_R}{dt}>0\right)$, and a left edge-state which is pumped down $\left(\frac{dE_L}{dt}<0\right)$.

\section{The QWZ Model: Chern Insulator in 2 Dimensions}\label{qwzmodel}

We note that the 2D Brillouin Zone $(k_x,k_y)$ is a torus, as is the 2D parameter space $(k,\lambda)$ for the adiabatic charge pump. Making the replacement $k \rightarrow k_x,\lambda(t) \rightarrow k_y$, we can {\it construct} a model for a 2D Chern Insulator out of a 1D charge pump, through a procedure called {\it dimensional extension} \cite{P6}.

The Chern number $n$ provides a link between the quantised Hall conductance $\sigma_{xy}$ and the change in polarisation $\Delta P = ne$.

On making a unitary rotation of our axes, $\sigma_x, \sigma_y, \sigma_z \rightarrow \sigma_z, \sigma_x, \sigma_y$ we arrive at the Qi-Wu-Zhang(QWZ) Hamiltonian:

\begin{eqnarray}\label{QWZ}
 H_{QWZ} ({\bf k}) &=& \lambda \sin(k_x a)\sigma_x+\lambda \sin(k_y a)\sigma_y \\ 
 \nonumber
 &+& (1+\lambda \cos (k_x a)+\lambda \cos(k_y a))\sigma_z 
\end{eqnarray}

The dispersion-relation can be read off from \ref{QWZ} as:

\begin{eqnarray}\label{disp}
 E_{\pm}({\bf k})&=&\pm [\lambda^2 \sin^2(k_xa)+\lambda^2 \sin^2(k_ya) \\ 
 \nonumber
 &+&(1+\lambda \cos(k_xa)+\lambda \cos(k_ya))^2)]^\frac{1}{2} 
\end{eqnarray}

The spectrum is generally gapped. However the gap closes at two specific values of $\lambda$:

\begin{enumerate}
 \item $\lambda=\frac{1}{2}$: the gap closes at $(k_x,k_y)=(\pm \pi,\pm \pi)$.
 \item $\lambda=-\frac{1}{2}$: the gap closes at $(k_x,k_y)=(0,0)$.
\end{enumerate}

As shown in Fig. 6(a) and 6(b) respectively.

\begin{figure}[h]
 \centering
 \includegraphics[width=0.9\columnwidth]{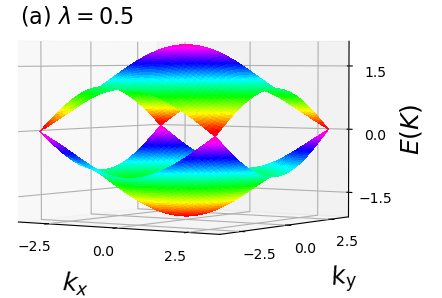}
 \includegraphics[width=0.9\columnwidth]{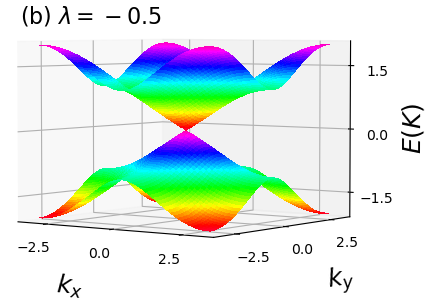}
 \caption{\small Gapless dispersion relations for the QWZ model: Dirac cones appear at (a) $(k_x,k_y) = (\pm \pi,\pm \pi)$ for $\lambda=0.5$ and (b) $(k_x,k_y)=(0,0)$ for $\lambda=-0.5$ }
\end{figure}

Near these points, the dispersion relation is approximately linear: they are called {\it Dirac points}.

\begin{figure}[h]
\centering
\begin{subfigure}{0.7\columnwidth}
\begin{tikzpicture}[scale=0.8]
\begin{axis}[xlabel = $d_x$, ylabel = $d_y$,zlabel=$d_z$, z buffer = sort,scale=0.8,yscale=1.5]
\addplot3[surf,samples = 15, samples y = 45,domain = 0:2*pi,domain y = 0:pi,thin](
            {sin(deg(\x))},
            {sin(deg(\y))},
            {1+cos(deg(\x))+cos(deg(\y))}
        );
\end{axis}
\draw [thick,red,dashed] (0,2.05)--(3.73,1.1);
\draw [thick,red,dashed] (1.89,0.8)--(1.89,2.4);
\end{tikzpicture}
\caption{$n=+1$}
\end{subfigure}
\begin{subfigure}{0.7\columnwidth}
\begin{tikzpicture}[scale=0.8]
\begin{axis}[xlabel = $d_x$, ylabel = $d_y$,zlabel=$d_z$, z buffer = sort,scale=0.8, yscale=1.5]
\addplot3[surf,samples = 15, samples y = 45,domain = 0:2*pi,domain y = 0:pi,thin](
            {sin(deg(\x))},
            {sin(deg(\y))},
            {-1+cos(deg(\x))+cos(deg(\y))}
        );
\end{axis}
\draw [thick,blue,dashed] (0,3.65)--(3.73,2.7);
\draw [thick,blue,dashed] (1.89,2.4)--(1.89,4);
\end{tikzpicture}
\caption{$n=-1$}
\end{subfigure}
\caption{\small Systems with $\lambda>\frac{1}{2}$ and $\lambda<-\frac{1}{2}$ have opposite topologies: For $\lambda>\frac{1}{2}$ (a) the Chern number $n$ is $+1$, while for $\lambda<-\frac{1}{2}$ (b), the orientation is reversed and $n=-1$}
\end{figure}
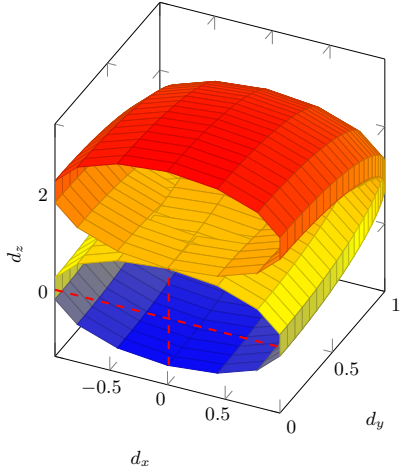
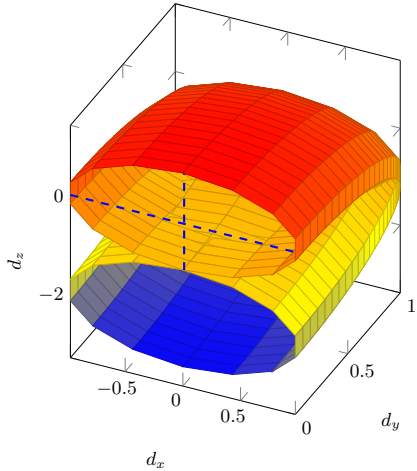

On writing $H({\bf k})$ as ${\bf d}({\bf k}).\vec \sigma$ we find that the parameters trace out a torus given by:

\begin{eqnarray*}
  d_x({\bf k})&=&\lambda \sin(k_xa)\\
  d_y({\bf k})&=&\lambda \sin(k_ya)\\
  d_z({\bf k})&=&1+\lambda \cos(k_xa)+\lambda \cos(k_ya)
\end{eqnarray*}

For $|\lambda|<\frac{1}{2}$ the torus entirely misses the origin and the system is topologically trivial.

For $|\lambda|>\frac{1}{2}$, we have a nonzero Chern Number:
\begin{enumerate}
 \item For $\lambda>\frac{1}{2}$, $n=1$: the lower part of the torus encloses the degeneracy (Fig. 7(a)).
 \item For $\lambda<-\frac{1}{2}$, $n=-1$: the orientation of the torus is reversed (Fig. 7(b)).
\end{enumerate}

Thus a topological phase transition takes place between $n=0$ and $n=1$ at $\lambda^+_c = \frac{1}{2}$, and another between $n=0$ and $n=-1$ at $\lambda^-_c=-\frac{1}{2}$. This is in fact expected from the dimensional extension of the Rice-Mele model. 

The Chern number determines the Hall conductance, as given by the Kubo formula $\sigma_{xy}=\frac{e^2}{2\pi\hbar} n$ \cite{P7}. Thus there is a jump from $\sigma_{xy}=0$ to $\sigma_{xy}=\pm\frac{e^2}{2\pi\hbar}$ across the topological phase transition.

Recall that the instantaneous left and right edge states of the Rice-Mele model are pumped down and up in energy respectively.

\begin{equation*}
 \frac{dE_L(t)}{dt}<0 \>\>\>\>\>\>\>\>\> \frac{dE_R(t)}{dt}>0
\end{equation*}

In the QWZ model with $\lambda>\frac{1}{2}$ the time variable $t$ has been mapped to the momentum variable $k_y$. This leads us to the relation:

\begin{equation*}
 \frac{dE_L(k)}{dk_y}<0 \>\>\>\>\>\>\>\>\> \frac{dE_R(k)}{dk_y}>0
\end{equation*}

This tells us that, for the QWZ model on open boundary conditions, the left and right edges {\it conduct in opposite directions}. These are called {\it chiral edge-modes}. 

So far, the appearance of chiral edge-modes in the QWZ model is just a prediction from dimensional extension. To explicitly demonstrate the existence of these edge-modes, we derive a low-energy effective Hamiltonian to model the behaviour near the $\lambda=-\frac{1}{2}$ Dirac point at $(0,0)$.

\begin{equation}\label{eff}
 H_{eff}({\bf k})=
 \begin{bmatrix}
  m & \lambda k_xa-i\lambda k_ya \\ 
  \lambda k_xa+i\lambda k_ya & -m
 \end{bmatrix}
\end{equation}

Where $m=1+2\lambda$: for the nontrivial phase ($n=-1$), $\lambda<-\frac{1}{2}$ and $m<0$ while for the trivial phase ($n=0$), $m>0$.

To detect the edge modes we allow a spatial variation in $m$ along the x-axis: $m(x)=-m$ for $x<0$ and $m(x)=+m$ for $x>0$. This creates an edge at $x=0$, separating the topologically nontrivial and trivial regions.

This system has translational invariance in the $y$-direction, but not in the $x$-direction. The wavefunction thus has the form $\psi_k(x,y) = e^{iky} \phi(x)$ such that:

\begin{equation*}
 \begin{bmatrix}
  m(x) & a\lambda(-i\frac{\partial}{\partial x}-ik_y) \\ 
  a\lambda(-i\frac{\partial}{\partial x}+ik_y) & -m(x)
 \end{bmatrix} 
 \phi(x) = \epsilon(k_y) \phi(x)
\end{equation*}

where $\epsilon(k_y)$ is the dispersion in the $y$-direction.

This equation has a solution localised at $x=0$ given by
\begin{equation}\label{jackreb}
\phi(x)=e^{\frac{1}{a}\int_0^x m(x')dx'}\begin{pmatrix}                                                                                       1\\                                                                                                     -i                                                                                                    \end{pmatrix}
\end{equation}

This eigenstate corresponds to the eigenvalue $\epsilon(k_y)=-a\lambda k_y$. Thus the group velocity of particles along the right edge is negative: $v_g^R=-a \lambda $. 
Similarly, for a system with $m>0$ for $x<0$ and $m<0$ for $x>0$, we have (on the left-edge) $v_g^L=+a \lambda $. 

These chiral edge-modes are oriented in a direction opposite to the predicted edge-modes from the dimensional extension of the Rice-Mele model. This is because, by focusing on $\lambda<-\frac{1}{2}$ we are working in the $n=-1$ sector rather than the $n=+1$ sector. The Chern number thus determines the orientation of the edge-modes as shown in Fig. 8.

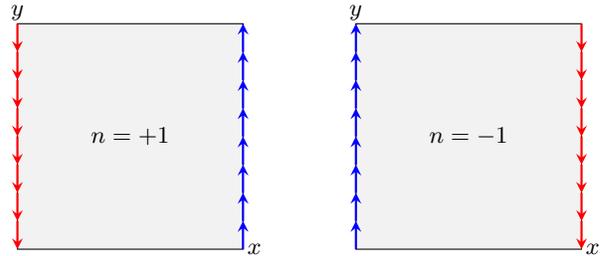
\begin{figure}[h]
\centering
\begin{tikzpicture}[scale=0.75]
\node at (6.2,0) {$x$};
\node at (2,4.2) {$y$};
\draw [fill=gray!10] (-4,4)rectangle (0,0);

\draw [red,thick][stealth-] (-4,3.5)--(-4,4);
\draw [red,thick][stealth-] (-4,3)--(-4,3.5);
\draw [red,thick][stealth-] (-4,2.5)--(-4,3);
\draw [red,thick][stealth-] (-4,2)--(-4,2.5);
\draw [red,thick][stealth-] (-4,1.5)--(-4,2);
\draw [red,thick][stealth-] (-4,1)--(-4,1.5);
\draw [red,thick][stealth-] (-4,0.5)--(-4,1);
\draw [red,thick][stealth-] (-4,0)--(-4,0.5);

\draw [blue,thick][stealth-] (0,4)--(0,3.5);
\draw [blue,thick][stealth-] (0,3.5)--(0,3);
\draw [blue,thick][stealth-] (0,3)--(0,2.5);
\draw [blue,thick][stealth-] (0,2.5)--(0,2);
\draw [blue,thick][stealth-] (0,2)--(0,1.5);
\draw [blue,thick][stealth-] (0,1.5)--(0,1);
\draw [blue,thick][stealth-] (0,1)--(0,0.5);
\draw [blue,thick][stealth-] (0,0.5)--(0,0);

\node at (0.2,0) {$x$};
\node at (-4.0,4.2) {$y$};
\draw [fill=gray!10] (2,4)rectangle (6,0);

\draw [red,thick][stealth-] (6,3.5)--(6,4);
\draw [red,thick][stealth-] (6,3)--(6,3.5);
\draw [red,thick][stealth-] (6,2.5)--(6,3);
\draw [red,thick][stealth-] (6,2)--(6,2.5);
\draw [red,thick][stealth-] (6,1.5)--(6,2);
\draw [red,thick][stealth-] (6,1)--(6,1.5);
\draw [red,thick][stealth-] (6,0.5)--(6,1);
\draw [red,thick][stealth-] (6,0)--(6,0.5);

\draw [blue,thick][stealth-] (2,4)--(2,3.5);
\draw [blue,thick][stealth-] (2,3.5)--(2,3);
\draw [blue,thick][stealth-] (2,3)--(2,2.5);
\draw [blue,thick][stealth-] (2,2.5)--(2,2);
\draw [blue,thick][stealth-] (2,2)--(2,1.5);
\draw [blue,thick][stealth-] (2,1.5)--(2,1);
\draw [blue,thick][stealth-] (2,1)--(2,0.5);
\draw [blue,thick][stealth-] (2,0.5)--(2,0);
\node at (-2,2) {$n=+1$};
\node at (4,2) {$n=-1$};
\end{tikzpicture}
\caption{\small The Chern number determines the orientation of the chiral edge-modes: $n=+1$ corresponds to counterclockwise modes while for $n=-1$ the modes are oriented clockwise}
\end{figure}

The solution \ref{jackreb} for $\phi(x)$ is called the {\it Jackiw-Rebbi solution} for the 2-dimensional Dirac equation, and crucially, is valid for any configuration $m(x)$ as long as $m(x)<0$ for $x<0$ and $m(x)>0$ for $x>0$. Since the sign of $m$ is a topological invariant, the chiral edge-modes are topologically protected, and hence {\it robust against local perturbations}.

The Chern number thus provides a link between the topology of the bulk Hamiltonian and the physics at the edge of the system. This is an example of the bulk-boundary correspondence \cite{P3}. 

\section{A Mechanism for the Topological Phase Transition}\label{mechanism}

In this section, we describe a physical {\it mechanism} for the onset of the topological phase transition using the notion of Time-Reversal Symmetry breaking.  

The Chern number on the 2D Brillouin Zone can be written as:

\begin{equation*}
 n = \frac{1}{4\pi} \int_{BZ} dS \>  F({\bf k})
\end{equation*}

Where $F(\bf k)$ is the {\it Berry curvature} given by: 

\begin{equation}
F({\bf k})=i\bra{\grad_{\bf k}u_{\bf k}}\cross\ket{\grad_{\bf k}u_{\bf k}}
\end{equation}

with $\partial_\alpha = \frac{\partial}{\partial k_\alpha}$. 

Under time-reversal $F({\bf k}) $ goes to $-F(-{\bf k})$, (see \cite{P3} for a detailed argument). Thus for a system with time-reversal symmetry $F({\bf k}) = - F(-{\bf k})$, ensuring that the Chern number associated with such a system vanishes: $n=0$. 

Thus, in order to have a topologically nontrivial phase, it is necessary to break time-reversal symmetry. 

However, {\it breaking time-reversal symmetry is not enough}.
Suppose we start out from a system in the flat-band limit: The system only has an onsite potential $H_0 = \sum_{i,j} (a^\dagger_{i,j}a_{i,j} - b^\dagger_{i,j}b_{i,j})$ (see Fig. 9(a)). Clearly $H_0$ does not break time-reversal symmetry. 

\begin{figure}[h]
\centering
 \includegraphics[width=0.9\columnwidth]{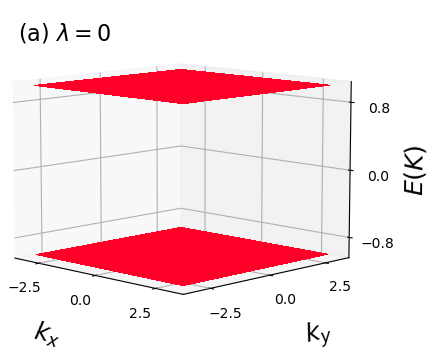}
 \includegraphics[width=0.9\columnwidth]{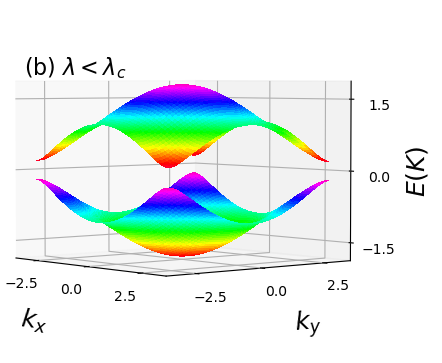}
\caption{\small Adding a time-reversal breaking term $\lambda H_1$ to the flat-band insulator (a) leads the bands to acquire a dispersion (b). For $\lambda<\lambda_c$ this is still a trivial insulator: The Chern number remains fixed at $n=0$ .}
\end{figure}

Now, we add a term $\lambda H_1$ which breaks time-reversal symmetry. Starting from $\lambda = 0$ and tuning $\lambda$ continuously, the bands acquire a dispersion (see Fig. 9(b)). However, it is impossible for the Chern number to {\it change continuously}, as it must remain fixed at an integer value. 

Now if the term $\lambda H_1$ leads the bulk gap to close at a certain value $\lambda=\lambda_c$, it is {\it impossible to adiabatically connect} systems with $\lambda>\lambda_c$ to systems with $\lambda<\lambda_c$. The system thus enters a different topological class, acquires a nonzero Chern number, and enters the topologically nontrivial phase. This phenomenon is called {\it band inversion}.

Trivial insulators ($\lambda<\lambda_c$) can be adiabatically connected to the flat-band limit while topological insulators ($\lambda>\lambda_c$) cannot.

Now we shall demonstrate the phenomenon of band inversion in the QWZ model. The momentum-space QWZ Hamiltonian may be written in the $(a,b)$ sublattice basis as:

\begin{eqnarray}
\nonumber
 H ({\bf k}) &=& \lambda e^{-ik_x a} \frac{\sigma_z+i\sigma_x}{2} + hc \\ 
 &+& \lambda e^{-ik_y a} \frac{\sigma_z+i\sigma_y}{2} + hc \\
 \nonumber
 &+& \sigma_z 
\end{eqnarray}

This allows us to come up with a {\it real-space description} of the QWZ model: The model arises from the tight-binding approximation for hopping a square lattice, with two orbitals $a$ and $b$ per lattice site.

The real-space Hamiltonian may be written as:
\begin{equation}\label{QWZreal}
 H_{QWZ} = H_0 + \lambda H_1 + \lambda H_2
\end{equation}

Here $H_0$ is an onsite potential:

\begin{equation*}
H_0 = \sum_{i=1}^{N} \sum_{j=1}^{N} \left(a^\dagger_{i,j}a_{i,j} - b^\dagger_{i,j}b_{i,j}\right)
\end{equation*}

$H_1$ is a standard nearest neighbour hopping term involving the same orbital on neighbouring sites:

\begin{eqnarray*}
 H_1 &=& \frac{1}{2} \sum_{i=1}^{N-1} \sum_{j=1}^{N} \left(a^\dagger_{i+1,j}a_{i,j} - b^\dagger_{i+1,j}b_{i,j} + hc \right)\\
 &+& \frac{1}{2} \sum_{i=1}^{N} \sum_{j=1}^{N-1} \left(a^\dagger_{i,j+1}a_{i,j} - b^\dagger_{i,j+1}b_{i,j} + hc\right)
\end{eqnarray*}

Finally $H_2$ is a term mixing $a$ and $b$ orbitals on nearest-neighbours:

\begin{eqnarray*}
 H_2 &=& \frac{1}{2} \sum_{i=1}^{N-1} \sum_{j=1}^{N} \left(ia^\dagger_{i+1,j}b_{i,j} + ib^\dagger_{i+1,j}a_{i,j} + hc \right) \\
 &+& \frac{1}{2} \sum_{i=1}^{N} \sum_{j=1}^{N-1} \left(a^\dagger_{i,j+1}b_{i,j} - b^\dagger_{i,j+1}a_{i,j} + hc\right)
\end{eqnarray*}

Note that the hopping is asymmetric in the $x$ and $y$ directions. More importantly for us, the hopping along the $x$ direction has a complex amplitude, {\it explicitly} breaking time-reversal symmetry.

$\lambda=0$ describes the {\it flat-band} limit ($H=H_0$) of the QWZ model. Time-reversal symmetry is preserved, and the Chern number thus vanishes: $n=0$. 

\begin{figure}[h]
\centering
 \includegraphics[width=0.9\columnwidth]{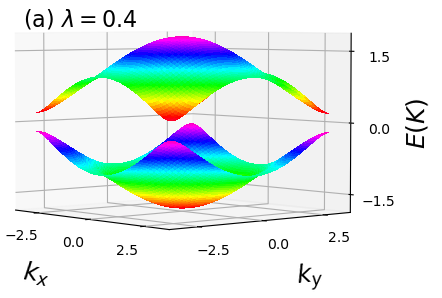}
 \includegraphics[width=0.9\columnwidth]{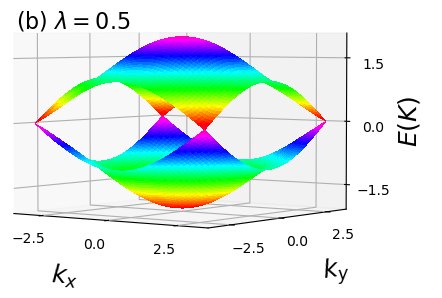}
 \includegraphics[width=0.9\columnwidth]{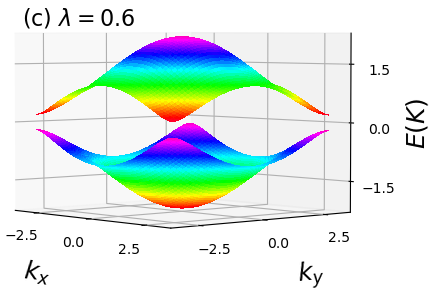}
 \caption{\small Topological phase transition in the QWZ model: for $\lambda<\frac{1}{2}$ (a) the system has a trivial gap. The system becomes metallic at $\lambda=\frac{1}{2}$ (b) and for $\lambda>\frac{1}{2}$ (c) the gap becomes nontrivial.}
\end{figure}
 
On tuning $\lambda$ adiabatically between $0$ and $\frac{1}{2}$, the bands acquire a dispersion (see Fig. 10(a)) there is no change in the Chern number, and the system remains in the trivial phase. 
 
At $\lambda=\frac{1}{2}$ the gap in the Bulk-Dispersion closes at $(k_x,k_y)=(\pm \pi,\pm \pi)$ (see Fig. 10(b)). 

For parameter values $\lambda>\frac{1}{2}$, the system enters a nontrivial topological phase with a jump in Chern number from $n=0$ to $n=+1$ (see Fig. 10(c)), leading to the appearance of chiral edge-modes and a jump in the Hall conductance. This phase cannot be adiabatically connected to the phase with $\lambda<\frac{1}{2}$.

The topology of the system is thus determined by a {\it competition} between the onsite term $H_0$ and the hopping term $H_2$ which breaks time-reversal symmetry.

Note that the dispersion relations at $\lambda=0.4$ and $\lambda=0.6$ appear similar. The difference between the phases is not in the bulk physics but in the {\it topological effects which appear at the edge}. 

In order to study the formation of edge-modes, we apply periodic boundary conditions in the $y$-direction and open boundary conditions in the $x$-direction. This allows us to separate the full Hamiltonian $H_{QWZ}$ into Hamiltonians describing 1-dimensional lattices, each for a different value of $k_y$.

\begin{equation}
 H_{QWZ} = \sum_{k_y} H_{1D}(k_y)
\end{equation}

On carrying out the partial fourier transform, we note that the hopping terms along the $y$-direction of the form $a^\dagger_{i,j+1}a_{i,j} + hc$ turn into onsite terms ($a^\dagger_i(k_y)a_i(k_y) \cos(k_ya)$) for the 1D model. Dropping the $k_y$ dependence of the operators, we write:

\begin{equation*}
 H_{1D} = H_0 + H_1 + H_2 + H_3
\end{equation*}

Where $H_0$ is an onsite term:
\begin{equation*}
 H_0 = \sum_{i=1}^N (1+\lambda \cos(k_ya)) (a^\dagger_ia_i-b^\dagger_ib_i)
\end{equation*}

$H_1$ mixes $a$ and $b$ orbitals at the same site:
\begin{equation*}
 H_1 = \sum_{i=1}^N i\lambda \sin(k_ya) (b^\dagger_ia_i-a^\dagger_ib_i)
\end{equation*}

$H_2$ is a standard hopping term:

\begin{equation*}
 H_2 = \sum_{i=1}^{N-1} \frac{\lambda}{2} (a^\dagger_{i+1}a_i-b^\dagger_{i+1}b_i) + hc
\end{equation*}

and $H_3$ is a hopping term with imaginary amplitude which connects the two different orbitals the orbital index: 
\begin{equation*}
 H_3 = \sum_{i=1}^{N-1} i\frac{\lambda}{2} (b^\dagger_{i+1}a_i+a^\dagger_{i+1}b_i) + hc
\end{equation*}

By placing the one-dimensional Hamiltonian $H_{1D}$ on a finite lattice (20 sites), we have solved the eigenvalue problem numerically, and demonstrated the formation of edge modes for $\lambda>\lambda_c$ (see Fig. 11).

\begin{figure}[h]
\includegraphics[width=\columnwidth]{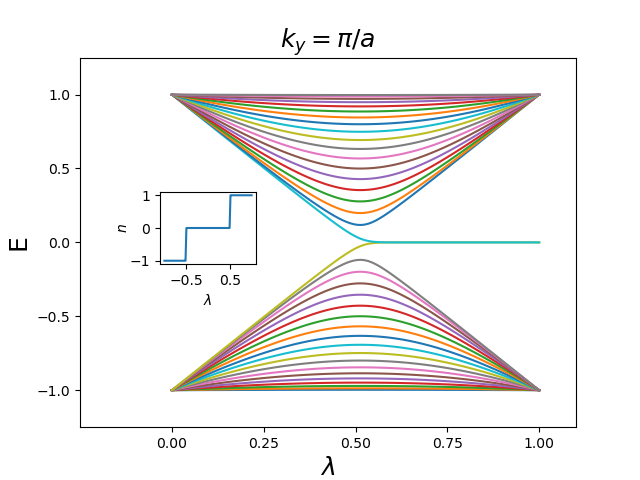}
\caption{\small Energy Levels for $H_{1D}$ with $k_y = \frac{\pi}{a}$: Zero-energy edge modes appear at $\lambda = \frac{1}{2}$. Inset: Phase diagram for the QWZ model.}
\end{figure}

As predicted, the edge-modes appear as $\lambda$ crosses the critical value $\lambda_c=\frac{1}{2}$. As $\lambda$ is increased further the and bulk gap opens, the edge-modes remain gapless.

To look for the formation of zero-energy edge-modes we look near the Dirac point at $k_y=\pi$ ($k_y=\pi + q_y$), with the system in the nontrivial phase $\lambda>\frac{1}{2}$. 

Choosing $\lambda=1$ allows us to neglect the onsite term $H_0$ to first order in $q_y a$. While the prefactor in $H_1$ is given by $-iq_ya$. 

Consider the state $\ket{R} = \frac{1}{\sqrt{2}}(\ket{a_N}-i\ket{b_N})$ localised at the right edge of the system. This state is created by the action of the operator $\Psi_R^\dagger = a^\dagger_N - i b^\dagger_N$ on the vacuum state. Defining $H_{hop} = H_2+H_3$, we find that:

\begin{eqnarray*}
 [H_{hop},\Psi_R^\dagger] &=& (a^\dagger_{N-1}-ib^\dagger_{N-1})(n_{b_N}-n_{a_N})\\
 &+&(a^\dagger_{N-1}-ib^\dagger_{N-1})(n_{a_N}-n_{b_N}) = 0
\end{eqnarray*}

$\ket{R}$ is thus a zero-energy eigenstate of $H_{hop}$. $\ket{R}$ is also an eigenstate of $H_1$:

\begin{equation*}
 H_1\ket{R}= q_ya \ket{R}
\end{equation*}

We thus have $H_{1D}\ket{R} =  q_ya \ket{R}$. Similarly, we have the eigenstate $\ket{L}=\frac{1}{\sqrt{2}}(\ket{a_1}+i\ket{b_1})$ localised at the left edge of the system: $H_{1D}\ket{L} = -q_ya \ket{L}$. 

The left and right edge-modes thus conduct in opposite directions, just as the Jackiw-Rebbi edge-modes described earlier. The Hamiltonian $H_{1D}$ allows us to explore topological effects in a finite-size system which are inaccessible via bulk-calculations.

These exactly-localised chiral edge-modes are also found in our numerical solution (see Fig. 12).

\begin{figure}[h]
 \centering
\includegraphics[width=\columnwidth]{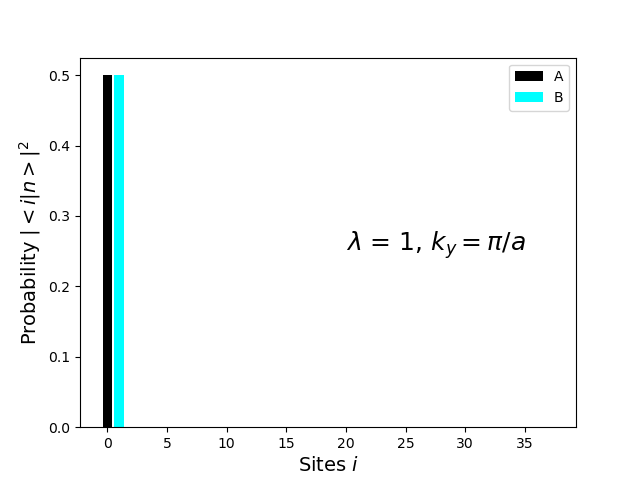}
\caption{\small Exactly localised edge-modes appear at $\lambda=1$ for $k_y = \frac{\pi}{a}$. These modes are not eigenstates of the onsite term $H_0$, which leads to the delocalisation of the edge-states over several lattice sites.}
\end{figure}

Moving away from this limit causes the edge-modes to {\it delocalise}: $\ket{R}$ and $\ket{L}$ are simultaneous eigenstates of $H_1$ and $H_{hop}$ but not of the onsite term $H_0$.

\section{Conclusion}\label{conc}
We analysed the formation of edge-modes in the QWZ model for a two-dimensional Chern insulator through the study of edge-states in one-dimensional chains, such as the RM model studied in Section \ref{rmmodel} and the 1D model $H_{1D}$ introduced in Section \ref{mechanism}.

We find that using simplified toy models, such as the toy sequence for the charge pump discussed in Section \ref{rmmodel}, allows us to accurately predict the topological features of more complicated systems which belong to the same topological class. 

Using the technique of {\it dimensional extension}, we constructed the Bulk Hamiltonian for the 2D QWZ model for a Chern insulator from the 1D RM charge pump, and demostrated that the topological phase transition in the RM model can be mapped to a corresponding topological phase transition in the QWZ model. The dynamics of edge-states in the RM model provided us an insight into the orientation of chiral edge-modes in the QWZ model.

Analysing the real-space properties of the QWZ model as done in Section \ref{mechanism} allowed us to come up with a physical mechanism for the onset of the topological phase transition. This analysis explicitly demonstrates the necessity of time-reversal symmetry breaking in bringing about a topological phase transition to the nontrivial phase. The real-space version of the model also allows us to solve the system on open-boundary conditions and examine the localisation properties of the edge-modes.

The 1D model presented in Section \ref{mechanism} offers an avenue for further exploration using both numerical and analytical methods.

Note: All figures, except graphs, have been have been created using the tikz package. \url{https://tikz.dev/}
\end{document}